\documentstyle[aps,prl,preprint]{revtex}

\newcommand{\BE}{\begin{equation}}
\newcommand{\EE}{\end{equation}}
\newcommand{\BA}{\begin{eqnarray}}
\newcommand{\EA}{\end{eqnarray}}

\begin{document}
\draft

\title{A Truly Minimal Left-Right Symmetric Extension\\ of the Standard Model}

\author{Fabio Siringo}
\address{Dipartimento di Fisica e Astronomia, 
Universit\`a di Catania,\\
INFN Sez. di Catania and INFM UdR di Catania,\\
Via S.Sofia 64, 95123 Catania, Italy}
\date{\today}
\maketitle
\begin{abstract}
By invoking the existence of a general custodial $O(2)$ symmetry, a
minimal Left-Right symmetric model based on the gauge group
$G=SU(2)_L\otimes SU(2)_R\otimes U(1)_{B-L}$
is shown to require the existence of only two physical Higgs bosons.
The lighter Higgs is predicted to have a small mass which could be
evaluated by standard perturbation theory. The fermionic mass matrices
are recovered by insertion of {\it ad hoc} fermion-Higgs interactions.
The model is shown to be undistinguishable from the standard model at
the currently reachable energies.
\end{abstract}
\pacs{PACS: 12.60.-i, 12.60.Fr, 12.60.Cn, 14.80.Cp
E-mail: fabio.siringo@ct.infn.it  Fax:++39 095 378 5231}

Left-Right (LR) symmetric extensions of the Standard Model (SM) 
have been extensively
studied since 1974 when they were first discussed\cite{pati,georgi}.
Some years ago the quantization and renormalization have been worked out by 
Duka et al.\cite{duka} who also gave an extended literature on the subject.
More recently radiative corrections have been considered in order to discuss the
phenomenological predictions of such models\cite{czakon}. 

LR models are based on the gauge group $G$
\BE
G=SU(2)_L\otimes SU(2)_R\otimes U(1)_{B-L}
\label{group}
\EE
and have the remarkable merit of predicting a phenomenology which is basically
undistinguishable from the standard model at the currently reachable energies.
However for the same reason the lack of any experimental evidence makes the choice of
the symmetry group - namely LR versus standard - a pure aesthetic matter.
From this point of view LR models have quite some advantages, as the observed L-R
asymmetry is explained by low energy breaking of the symmetry, and does not require
to be inserted by hand as it is the case for the standard model. Moreover the $U(1)$
generator gets a physical interpretation as the $B-L$ number.
On the other hand in their ``minimal'' version LR symmetric models require the existence
of many new particles: the most disturbing are the ten physical Higgs particles required
(4 charged and 6 real neutral bosons) to be compared with one only physical Higgs boson
predicted by the SM. 

In this paper it is shown that the proliferation of Higgs particles is not necessary
and can be avoided in a truly minimal version of L-R symmetric model.

Any viable gauge model for electro-weak interactions must give an answer to two 
quite different problems: (i) the breaking of symmetry - from the full gauge group 
to the electro-magnetic abelian group $U(1)_{em}$ - 
which gives a mass to the gauge bosons
and thus explains the known structure of weak interactions; (ii) the mass matrices
for fermions. Even in the standard model these two problems are addressed in different 
independent steps, and with a different degree of success. In fact, while the first
problem is given a full and satisfactory solution, the mass generation for
fermions is only described by {\it ad hoc}  insertion of a handful of coupling 
parameters. Thus the mass hierarchy problem has not found any genuine solution.
Moreover this second aspect is strongly related to the nature of the Higgs sector
which is still unexplored from the experimental point of view. In LR models, the
proliferation of Higgs particles is a direct consequence of the very complicated
structure of the Higgs sector, required in order to explain the spontaneous 
breaking of L-R symmetry. Unless the Higgs bosons are composite objects, as
predicted by top\cite{top} or neutrino\cite{neutrino} condensation models, the
prospect of such a large number of elementary fields is not very attractive.

By invoking the existence of a general custodial symmetry, we show that a truly
minimal LR model, based on the gauge group $G$, 
only requires the existence of two physical neutral scalar Higgs
bosons in order to address point (i). Point (ii) remains open to several quite
different descriptions which are compatible with the proposed symmetry breaking path.
{\it Ad hoc} insertion of free coupling parameters
between fermions and the two Higgs fields still allows for a full description of the 
fermion mass matrices, at the extra cost of inserting 
non-renormalizable terms in the lagrangian. 
Of course this choice would not give any real answer to the 
mass hierarchy problem which remains unsolved.  
 
The LR symmetric lagrangian is the sum of a fermionic term ${\cal L}_f$, 
a Yang-Mills term for the gauge bosons ${\cal L}_{YM}$, a Higgs term 
${\cal L}_H$ and eventually the Higgs-fermions interaction term
${\cal L}_{int}$.

In order to deal with point (i) above, we need to specify ${\cal L}_H$:
\BE
{\cal L}_H=-{1\over 2} \vert D_L^\mu \chi_L\vert^2
-{1\over 2} \vert D_R^\mu \chi_R\vert^2+V(\chi_L,\chi_R)
\label{Lhiggs}
\EE
where the covariant derivative $D_a^\mu$ is defined according to
\BE
D_a^\mu=\left(\partial^\mu-i g_a \vec A^\mu_a \vec T_a
+i\tilde g B^\mu {Y\over 2}\right), \qquad\qquad a=L,R.
\label{covariant}
\EE
$\vec T_L$, $\vec T_R$ and $Y$ are the generators of $SU(2)_L$, $SU(2)_R$
and $U(1)_{B-L}$ respectively, with couplings $g_L=g_R=g$ and $\tilde g$.
As usual the electric charge is given by $Q={T_L}_3+{T_R}_3+Y/2$.
The Higgs fields $\chi_a$ are doublets 
\BE
\chi_L={\chi^+_L \choose\chi^0_L},\qquad \chi_R={\chi^+_R \choose\chi^0_R}
\label{higgs}
\EE
with the trasformation properties
\BE
\chi_L\equiv(2,1,1),\qquad\qquad \chi_R\equiv(1,2,1)
\EE
A standard ${\cal L}_{YM}$ is considered for the seven gauge fields $\vec A^\mu_L$,
$\vec A^\mu_R$ and $B^\mu$.
Fermions are described by doublets of spinors $\psi_L$, $\psi_R$ with the
transformation properties 
\BE
\psi_L\equiv(2,1,B-L),\qquad\qquad \psi_R\equiv(1,2,B-L).
\EE
Their lagrangian term ${\cal L}_f$ follows
\BE
{\cal L}_f=-\bar\psi_L \gamma_\mu D_L^\mu\psi_L
-\bar\psi_R \gamma_\mu D_R^\mu\psi_R
\label{Lf}
\EE

The lagrangian ${\cal L}={\cal L}_f+{\cal L}_{YM}+{\cal L}_H$ 
is fully symmetric for L-R exchange. Moreover we notice that
for a vanishing coupling $g,\tilde g\to 0$, and neglecting the
contribution of $V(\chi_L,\chi_R)$, the free part of 
${\cal L}_f+{\cal L}_H$ has a global O(2) symmetry, as it is 
invariant for rotations in the $L-R$ plane. We may define doublets
of doublets according to
\BE
\Phi={\chi_L \choose\chi_R},\qquad \Psi={\psi_L \choose\psi_R}
\label{doublets}
\EE
and the free lagrangian reads
\BE
{\cal L}_f+{\cal L}_H=-\vert \partial_\mu \Phi\vert^2
-\bar\Psi\gamma^\mu\partial_\mu\Psi
\label{freeL}
\EE
which is invariant for $\Phi\to\Phi^\prime=R(\theta)\Phi$ and
$\Psi\to\Psi^\prime=R(\theta)\Psi$, where $R(\theta)\in O(2)$
is the $2\times 2$ rotation matrix of angle $\theta$.
While the physical meaning of this continuous global symmetry is not
evident, we may assume that the full lagrangian should be $O(2)$ 
invariant in the limit $g,\tilde g\to 0$ of no gauge coupling.
According to such assumption the Higgs potential $V(\chi_L,\chi_R)$
should be invariant for rotations in the L-R plane.
This custodial symmetry makes the potential $V$ a function of the
rotational invariant field $\rho$
\BE
\rho^2=\chi_L^2+\chi_R^2,
\label{rho}
\EE
and this is going to hold even in presence of
finite gauge interactions which break the O(2) symmetry.
Provided that $V(\rho)$ has a minimum for a non-zero expectation value
of $\rho$, according to (\ref{rho})
the minimum is going to be on a circle in the
$\chi_L$,$\chi_R$ plane. For the real vacuum, the actual value of 
$\theta$ is only determined by chance. Thus the custodial symmetry
would give a natural path towards the L-R symmetry breaking.

If the gauge interactions are allowed to break the O(2) symmetry then
infinite renormalizations of the O(2) breaking terms would spoil the
symmetry of the scalar potential. Thus we must assume that the quartic
terms in the potential are set to zero by some unspecified physics in
the high energy theory.

The physical content of the theory becomes more evident in unitarity
gauge. According we set $\chi^+_a=0$ and choose $\chi^0_a$ real.
The covariant derivative of $\chi_a$ reads
\BE
D^\mu\chi_a={-i{g\over\sqrt{2}} {W_a^-}^\mu\chi^0_a\choose
(i{g\over 2} {A_a}_3^\mu+i{{\tilde g}\over 2}B^\mu+\partial^\mu)\chi_a^0}
\label{derivative}
\EE
where ${W_a^\pm}={1\over\sqrt{2}}({A_a}_1\pm i{A_a}_2)$.
For a non-zero expectation value of $\rho$ we get the following
vacuum expectation values for the Higgs fields:
\BE
\langle\chi^0_L\rangle=v=\rho\sin\theta;\qquad 
\langle\chi^0_R\rangle=w=\rho\cos\theta
\label{vev}
\EE
and we assume that by chance $\theta$ is very small ($v<<w$). 
Insertion of Eq.(\ref{derivative}) in the lagrangian (\ref{Lhiggs}) yields
the mass matrix for the gauge bosons.

The charged $W^\pm_L$ and $W^\pm_R$ are decoupled with masses
\BE
M_{W(L)}={{gv}\over 2}, \qquad
M_{W(R)}={{gw}\over 2}.
\label{MW}
\EE
Thus the angle $\theta$ determines the mass ratio $\tan\theta=M_{W(L)}/M_{W(R)}$.
For the neutral gauge bosons $B^\mu$, ${A_L}_3^\mu$ and ${A_R}_3^\mu$ we get
the mass matrix $M^2$
\BE
M^2={1\over 4}\left(\matrix{
\tilde g^2(v^2+w^2) &   g\tilde gv^2  & g\tilde gw^2 \cr
g\tilde gv^2   &    g^2v^2     &   0    \cr
g\tilde gw^2   &     0       & g^2w^2   \cr
}\right).
\label{mass}
\EE
There is a vanishing eigenvalue for the electromagnetic unbroken $U(1)$ eigenvector
\BE
A^\mu={e\over g} {A_L}_3^\mu+{e\over g} {A_R}_3^\mu-{e\over {\tilde g}} B^\mu
\label{A}
\EE
while the non-vanishing eigenvalues are given by a small value
\BE
M_Z^2={{g^2v^2(g^2+2\tilde g^2)}\over{g^2+\tilde g^2}}+{\cal O}(v^2/w^2)
\label{MZ}
\EE
and a large one
\BE
M_{Z^\prime}^2=\left(M_{W(L)}^2+M_{W(R)}^2\right)(1+\tilde g^2/g^2)-M_Z^2
\label{MZprime}.
\EE
The corresponding eigenvectors are
\BE
\left(\matrix{ A \cr Z \cr Z^\prime \cr}\right)=
\left(\matrix{
-e/{\tilde g} &  e/g & e/g\cr
g/D_L & \displaystyle
{{\tilde g \cos^2 \theta_L}\over{D_L\sin^2\theta_L}} &\displaystyle
{{\tilde g \cos^2 \theta_L}\over{D_L(\tan^2\theta-\cos^2\theta_L)}} \cr
g/D_R & \displaystyle 
{{\tilde g \cos^2 \theta_R}\over{D_R(\tan^{-2}\theta-\cos^2\theta_R)}} &
\displaystyle 
{{\tilde g \cos^2 \theta_R}\over{D_R\sin^2\theta_R}} \cr}\right)
\left(\matrix{ B \cr {A_3}_L \cr {A_3}_R \cr}\right)
\label{eigenvector}
\EE

where
\BE
D_L=\sqrt{g^2+\tilde g^2\left(\tan^{-4}\theta_L+
\cos^4\theta_L/(\tan^2\theta-\cos^2\theta_L)^2\right)},
\EE
\BE
D_R=\sqrt{g^2+\tilde g^2\left(\tan^{-4}\theta_R+
\cos^4\theta_R/(\tan^{-2}\theta-\cos^2\theta_R)^2\right)}
\EE
and the Weinberg angles $\theta_L$ and $\theta_R$ are defined
according to $\cos\theta_L=M_{W(L)}/M_Z$,
$\cos\theta_R=M_{W(R)}/M_{Z^\prime}$.

The transformation matrix (\ref{eigenvector}) can be inverted,
and neglecting contributions of order ${\cal O} (v^2/w^2)$ and the
heavy $Z^\prime$  we find
\BE
B^\mu=-{e\over {\tilde g}}A^\mu
+{g\over {\tilde g}}\tan\theta_L\sin\theta_LZ^\mu+\dots
\label{B}
\EE
\BE
{A_3}_L^\mu={e\over g}A^\mu
+\cos\theta_LZ^\mu+\dots
\label{A3L}
\EE
\BE
{A_3}_R^\mu={e\over g}A^\mu
-\tan\theta_L\sin\theta_LZ^\mu+\dots
\label{A3R}
\EE
while the normalization condition ensures that $e^2=g^2\sin^2\theta_L$.

Thus the SM phenomenology is recovered with $\theta_L$ playing the role of the standard
Weinberg angle\cite{theorem}, and $v$ determined by the Fermi constant. 
In fact insertion of Eqs.(\ref{B}),(\ref{A3L}),(\ref{A3R}) in the fermion lagrangian
(\ref{Lf}) yields the standard model effective lagrangian up to ${\cal O}(v^2/w^2)$
corrections. Of course, at low energy, all the effects of the heavy $Z^\prime$ and
$W^{\pm}_R$ are suppressed. 

The Higgs sector is very simple, as we only have two neutral scalar fields
$\chi^0_L$, $\chi_R^0$. In the limit $g,\tilde g\to 0$ the mass matrix has a vanishing
eigenvalue in the point 
$\chi^0_L=v$, $\chi^0_R=w$. Thus the physical fields are a radial Higgs
$\rho=\sqrt{v^2+w^2}$ and a tangential zero-mass Goldstone boson. For the radial
field the mass is determined by the unknown potential $V(\rho)$ and, as for the SM,
there are only loose bounds on its value. The cost of LR symmetry breaking seems
to be the occurrence of a zero-mass Higgs field, however for finite gauge
couplings $g,\tilde g\not=0$ the $O(2)$ custodial symmetry is not an exact symmetry
of the full model, and the tangential ``would be'' Goldestone boson is expected to 
acquire a mass. 

Until now we have not addressed point (ii) (i.e. the origin of fermion mass matrices), 
and we avoided to discuss any Higgs-fermion interaction. 
In the SM the mass of fermions is recovered
by insertion of {\it ad hoc} interactions. Taking aside composite Higgs theories, which
would be compatible with the present LR minimal model, 
insertion of a fermion-Higgs interaction still remains
the simplest way to predict fermion mass matrices. In order to avoid the
proliferation of Higgs fields we may build up the four composite matrices 
$\chi_a\chi_b^\dagger$ with $a,b=L,R$. For each set of $ab$ labels 
$\chi_a\chi_b^\dagger$ is a matrix since 
both $\chi_a$ and $\chi_b$ are doublets according to
their definition (\ref{higgs}). Moreover, 
the $O(2)$ custodial symmetry requires that we regard 
$\Phi\Phi^\dagger\equiv \chi_a\chi_b^\dagger$ 
as a matrix of matrices. An $O(2)$ invariant interaction term may be written as
$\bar\Psi\Phi\Phi^\dagger\Psi$.
We may also define the adjoint doublets 
\BE
\tilde\chi_a={(\chi^0_a)^\star \choose (-\chi^+_a)^\star}
\qquad \tilde\Phi={\tilde\chi_L \choose \tilde\chi_R}
\label{adjoint}
\EE
that have the same transformation properties as $\chi_a$ and $\Phi$ respectively.
Thus the more general interaction, invariant for $G$ and $O(2)$ 
transformations, reads
\BE
{\cal L}_{int}=-\alpha_1 \bar\Psi\Phi\Phi^\dagger\Psi
-\alpha_2 \bar\Psi\tilde \Phi\tilde \Phi^\dagger\Psi
\label{interaction}
\EE
where the couplings $\alpha_1$, $\alpha_2$  change for different fermionic
doublets. If we require exact conservation of the leptonic number, then no
Majorana mass term is allowed, and neutrinos are regarded as standard
Dirac fermions.
The interaction term may be simplified by noticing that 
$\bar\psi_L\psi_L=\bar\psi_R\psi_R=0$. Thus we find
\BE
{\cal L}_{int}=-\alpha_1\bar\psi_L\chi_L\chi_R^\dagger\psi_R
-\alpha_2\bar\psi_L\tilde\chi_L\tilde\chi_R^\dagger\psi_R +h.c.
\EE
Then in unitarity gauge, by inserting
\BE
\chi_a={0 \choose \chi^0_a},
\qquad \tilde\chi_a={\chi^0_a \choose 0},
\qquad \psi_a={u_a \choose d_a}.
\label{unitarity}
\EE
the interaction term reads
\BE
{\cal L}_{int}=-\chi^0_L\chi^0_R\left(
\alpha_1\bar d_L d_R+\alpha_2\bar u_L u_R\right) +h.c.
\EE
At low energy the vacuum expectation values of the Higgs fields
give mass terms to the fermions. Up and down components of the
fermionic doublets get different masses $m_u$, $m_d$
\BE
m_u=\alpha_2 v w,\qquad m_d=\alpha_1 v w
\label{mf}
\EE
The generalization to the case of three fermionic flavours
is straightforward, with the couplings $\alpha_1$, $\alpha_2$
replaced by matrices. 

The cost we pay is the insertion of a non-renormalizable term in the
lagrangian. However the couplings $\alpha_i$ are very small as
\BE
\alpha_i=\left({{g^2}\over{4 M_{W(R)}}}\right)
\left({{m_i}\over{M_{W(L)}}}\right)
\EE
and scale as $1/M_{W(R)}$. 
The large mass of the top quark suggests that the cut-off scale cannot be much
bigger than $M_{W(R)}$, and thus the effective theory makes sense only to
energies a little higher than $M_{W(R)}$. However, if $\theta$ is very small,
that energy scale is considerably higher than any known mass. A full analysis
of the phenomenology of the non-renormalizable interactions is out of the aim
of the present paper. On the other hand, we stress that the existence of such
non-renormalizable terms in the lagrangian was not required in order to address
point (i) (symmetry breaking and the known structure of weak interactions) which
is the main aim of the paper. Inclusion of {\it ad hoc} terms like ${\cal L}_{int}$
is just the simplest way to reproduce the fermion mass matrices.

An open question is the mass of the ``would be'' tangential Goldstone boson.
In principle its value could be evaluated by standard perturbation theory.
In a simplified picture we may assume that zero-point energies would 
contribute\cite{coleman} a finite effective potential term
$V_b(M)=3/(64\pi^2) M^4\ln (M^2/\mu_b^2)$ for any vector boson field 
with mass $M$, and a term
$V_f(m)= -4/(64\pi^2)m^4\ln (m^2/\mu_f^2)$ for any fermionic field with mass $m$. 
The energies $\mu_b$, $\mu_f$ depend on the cut-off scale and can be regarded as
free parameters. According to eqs.(\ref{vev}),(\ref{MW}),(\ref{mf}), summing up
over all the fermionic masses $m_j=\alpha_jvw=\alpha_j\rho^2\sin\theta\cos\theta$
and over all the bosonic masses, and assuming $M_Z\approx M_{W(L)}=vg/2$,
$M_{Z^\prime}\approx M_{W(R)}=gw/2$, we obtain the following effective
potential contribution
\BA
V_{eff}(\theta)&=&{{\rho^4}\over {64\pi^2}}\left[
{{(3\cdot 3)g^4}\over {16}}\sin^4\theta
\ln\left({{g^2\rho^2\sin^2\theta}\over{4\mu^2_L}}\right)+\right\delimiter0\nonumber\\
&+&\left\delimiter0{{(3\cdot 3)g^4}\over {16}}\cos^4\theta
\ln\left({{g^2\rho^2\cos^2\theta}\over{4\mu^2_R}}\right)
-4\rho^4\sum_j \alpha_j^4\sin^4\theta\cos^4\theta
\ln\left({{\alpha_j^2\rho^4\sin^2\theta\cos^2\theta}\over{\mu_j^2}}\right)
\right]
\label{eff}
\EA
where the sum over $j$ runs over all the known fermions. The ratio 
$\rho^4\alpha_j^4/g^4\approx(m_j/M_{W(L)})^4$ is very small and negligible for almost
all fermions. Only the top quark contributes to $V_{eff}$ which reads
\BE
V_{eff}=const.\times\left[\sin^4\theta\ln(a_L\sin^2\theta)+
+\cos^4\theta\ln(a_R\cos^2\theta)-\xi\sin^4\theta\cos^4\theta\ln(b\sin^2\theta\cos^2\theta)\right]
\label{effpot}
\EE
where $a_L=(g\rho/2\mu_L)^2$, $a_R=(g\rho/2\mu_R)^2$, $b=\alpha_t^2\rho^4/\mu_t^2$
and 
\BE
\xi={4\over{9\cos^4\theta_0}}\left({{m_t}\over{M_{W(L)}}}\right)^4
\EE
with $\theta_0$ fixed at the phenomenological value of $\theta$.
Insertion of  $M_{W(L)}=81.5$~GeV, $m_t=181$~GeV, $\cos\theta_0\approx 1$
yields $\xi\approx 11.$ The diagram of $V_{eff}$ is reported in Fig.~1 for $a_L=a_R=b$
taken in the range from 0.65 (upper curve) to 1.1 (lower curve).
This effective potential term breaks the $O(2)$ invariance, and has an absolute
minimum for small values of $\theta$ in a broad range of parameters.
Detailed calculations are called for in order to compare the predictions of the
model with future experimental data on the Higgs mass.

In summary we have shown that the existence of a custodial $O(2)$ symmetry
would give a natural path towards the LR symmetry breaking without requiring
complex Higgs sectors. In its minimal version the model only requires two physical
neutral Higgs bosons, and predicts a phenomenology which is undistinguishable
from the SM at the currently reachable energies. Moreover a light Higgs field
is predicted whose mass could be evaluated by standard perturbative calculations.

\begin{figure}
\caption{Finite contributions of zero point energies to the effective potential according to
Eq.(\ref{effpot}) of the text. A constant has been subtracted in order to set $V_{eff}(0)=0$,
and an arbitrary scale factor has been inserted. The free parameters $a_L$, $a_R$ and $b$ are
taken to be $a_L=a_R=b=6.5,8.0,9.5,1.1$ (going from the upper to the lower curve respectively).
A minimum is present for a small value of the angle $\theta$
which decreases when the free parameters decrease.}
\end{figure}


\begin{thebibliography} {99}
\bibitem{theorem} That LR theories predict the SM phenomenology has been shown 
   on very general grounds by Senjanovic\cite{senjanovic} who employed the method of
   Georgi and Weinberg\cite{georgi2}.
\end{thebibliography}
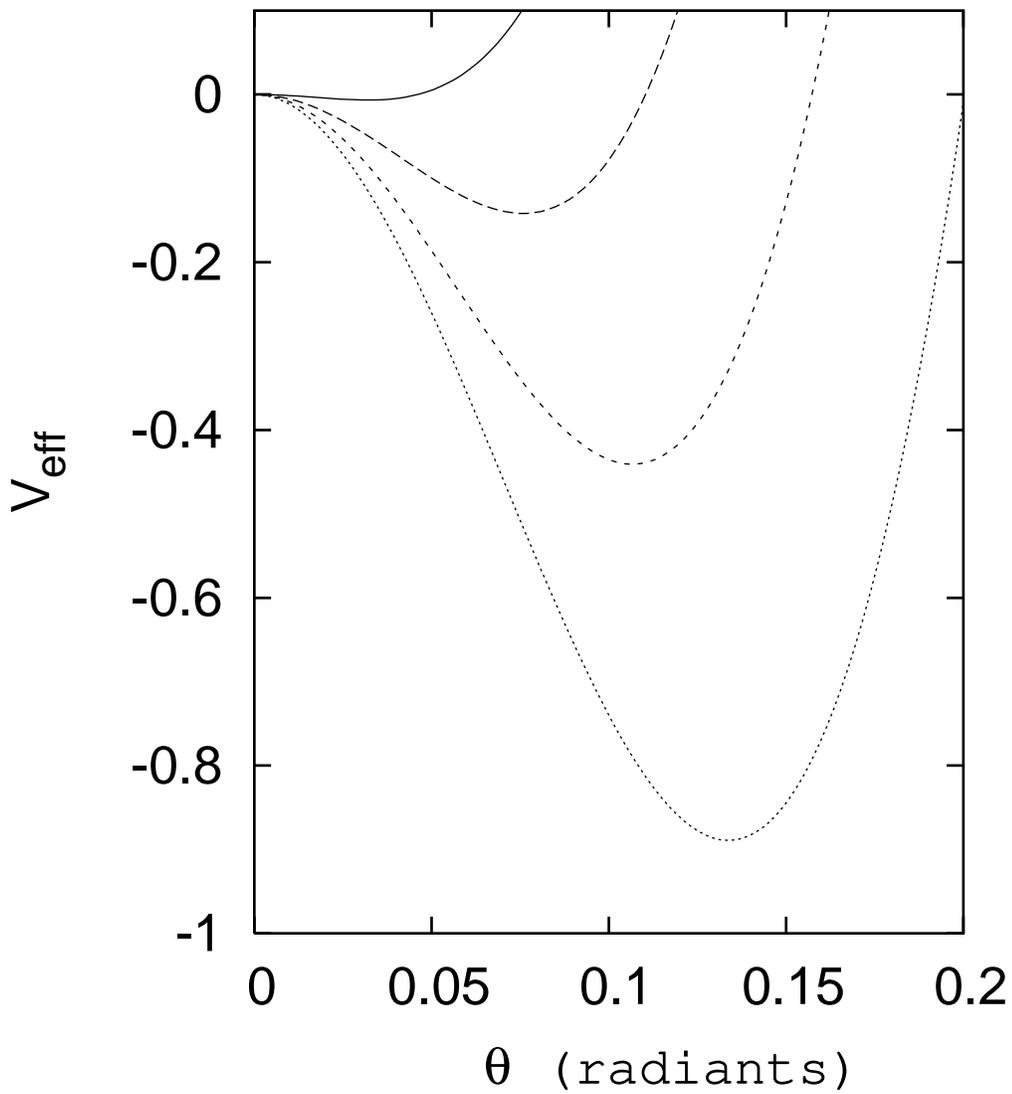
\end{document}